\RequirePackage[2020-02-02]{latexrelease}
\documentclass[journal]{IEEEtran}

\usepackage{url}
\usepackage{lscape}

\usepackage{booktabs}
\usepackage{longtable}
\usepackage{graphicx}
\usepackage{commath}
\usepackage{multirow}
\usepackage{epstopdf}
\usepackage{pdfpages}
\usepackage{dblfloatfix}
\usepackage[export]{adjustbox}
\usepackage[utf8]{inputenc}
\usepackage[printwatermark]{xwatermark}
\usepackage{xcolor,colortbl}

\usepackage{tikz}

\usepackage{algorithm}
\usepackage{academicons}
\usepackage{comment}
\usepackage{blindtext}
\usepackage{mathtools}
\usepackage{amssymb}
\usepackage{amsmath}
\usepackage{mathtools}
\usepackage{hyperref}
\usepackage{algpseudocode}
\usepackage[hang,flushmargin]{footmisc}
\usepackage{lineno,hyperref}
\usepackage{xcolor,colortbl}

\begin{document}

\title{Internet of Nano, Bio-Nano, Biodegradable and Ingestible Things: A Survey}

\author{Şeyda Şenturk, İbrahim Kök and
\and Fatmana Şentürk

\IEEEcompsocitemizethanks{
\IEEEcompsocthanksitem Ş. Şentürk is with the Department of Food Engineering, Pamukkale University, Denizli, TR e-mail:seydasenturk@gmail.com
\protect
\IEEEcompsocthanksitem İ. Kök and F. Şentürk are with the Department of Computer Engineering, Pamukkale University, Denizli,
	TR e-mail:ikok@pau.edu.tr, fatmanas@pau.edu.tr.}}

\maketitle             

\begin{abstract}
In recent years, advances in biotechnology, nanotechnology and materials science have led to development of revolutionizing applications in Internet of Things (IoT). In particular, the interconnection of nanomaterials, nanoimplants and nanobiosensors with existing IoT networks have inspired the concepts of Internet of Nano Things (IoNT),  Internet of Bio-Nano Things (IoBNT), Internet of Biodegradable Things (IoBDT) and Internet of Ingestible Things (IoIT). To date, although there are several survey papers that addressed these concepts separately, there is no current survey covering all studies in IoNT, IoBNT, IoBDT and IoIT. Therefore, in this paper, we provide a complete overview of all recent work in these four areas. Furthermore, we emphasize the research challenges, potential applications, and open research areas.
\end{abstract}

\section{Introduction}

The number of Internet usage and internet-based applications in our daily life is increasing day by day \cite{miraz2018internet}. This increase also brings about an exponential increase in the number of devices connected to the Internet. In this context, it is estimated that more than 75 billion devices will be connected to the Internet by the end of 2025 \cite{nawaratne2018self}. As a natural consequence of these developments, the Internet of Things (IoT) concept has become the focus of research and development, especially in the last 15 years \cite{fouad2020analyzing}. IoT concept represents the connection and communication of all kinds of physical things such as sensors, actuators, personal electronic devices in the real world via the Internet \cite{akyildiz2015internet}. Today, IoT applications appear in many areas such as smart transportation, real-time monitoring systems, smart cities, smart grids, smart environmental monitoring, medical and health systems, and smart buildings \cite{kok2020deepmdp}. On the other hand, thanks to the advancements in materials science and nanotechnology, IoT application areas are expanding further and paving the way for the formation of new concepts such as Internet of Nano Things (IoNT), Internet of Bio-Nano
Things (IoBNT) and Internet of Biodegradable Things (IoBDT).  

\begin{table*}[!htb]
\label{existing_survey}
\caption{Existing survey papers and relation to this paper}
\resizebox{\textwidth}{!}{%
\begin{tabular}{lllllll}
\toprule
Survey Paper & IoT & IoNT & IoBNT & IoBDT & IoIT & Description \\ \midrule
\cite{pramanik2020advancing} & $\times$ & \checkmark & $\times$ & $\times$ & $\times$ & \begin{tabular}[c]{@{}l@{}}This study discusses the usage scenarios, network architecture, communication paths, pros and \\ cons of IoNT and nano sensors in modern healthcare.\end{tabular} \\
\cite{cruz2019understanding} & $\times$ & \checkmark & $\times$ & $\times$ & $\times$ & \begin{tabular}[c]{@{}l@{}}This paper presents the features, network architecture, application possibilities, key challenges and\\ future trends of IoNT.\end{tabular} \\
\cite{ali2016internet} & $\times$ & \checkmark & $\times$ & $\times$ & $\times$ & \begin{tabular}[c]{@{}l@{}}This paper briefly summarizes the network models of IoNT and the difficulties encountered in its \\ implementation in healthcare.\end{tabular} \\
\cite{velvizhi2020biodegradable} & \checkmark & $\times$ & $\times$ & $\times$ & $\times$ & \begin{tabular}[c]{@{}l@{}}This paper focuses more on the digestibility of biodegradable fraction of solid waste. It addresses the \\ use of IoT technologies to reveal the problems and challenges faced in solid waste management.\end{tabular} \\
\cite{steiger2019ingestible} & $\times$ & $\times$ & $\times$ & $\times$ & \checkmark & \begin{tabular}[c]{@{}l@{}}This paper explores the safety, communication, power, tissue interactions of ingestible electronics and \\ reviews therapeutic interventions such as macromolecule delivery, drug delivery and distribution of \\ electrical signals.\end{tabular} \\
\cite{kalantar2017ingestible} & $\times$ & $\times$ & $\times$ & \checkmark & $\times$ & \begin{tabular}[c]{@{}l@{}}This article comprehensively details ingestible sensors and sensing capsules in terms of body physiology \\ and technology. It also explains the effects and applications on the intestinal structure.\end{tabular} \\
\cite{kuscu2021internet} & $\times$ & $\times$ & \checkmark & $\times$ & $\times$ & \begin{tabular}[c]{@{}l@{}}This paper presents the key components, applications, technological challenges and future research \\ directions of IoBNT\end{tabular} \\
Our Paper & \checkmark & \checkmark & \checkmark & \checkmark & \checkmark & \begin{tabular}[c]{@{}l@{}}Our paper covers all nanonetwork paradigms and potential application areas,  architectural models and \\ communication structures. It also comprehensively addresses current challenges and future research\end{tabular} \\ \hline
\end{tabular}
} 
\end{table*}
Although these concepts are just emerging today, the history of nanotechnology, which forms the basis of these concepts, goes back to the 1950s. In 1959, Richard Feynman first discussed the possibility of directly manipulating materials at the atomic scale and the idea of reproducing everything by miniaturization on a small scale in his famous talk entitled "\textit{There's Plenty of Room at the Bottom}" \cite{128057}. In 1974, the term "Nanotechnology" was first coined by Norio Taniguchi to describe dimensional accuracy \cite{hulla2015nanotechnology}. Nanotechnology refers to the study and application of tiny things measured at the atomic and molecular scale. It has the potential to produce many new technological materials and devices for the benefit of people by bringing together many disciplines such as engineering, medicine, biology, physics and chemistry \cite{pramanik2020advancing}. Nano things are devices ranging from 1 to 100 nanometer (nm) and can perform tasks such as collecting, creating, computing, processing and transmitting data at nanoscale \cite{nayyar2017internet,al2020cognitive}. Today, it has revealed promising networking paradigms such as IoNT, IoBNT, IoBDT, and IoIT, which are formed by interconnecting nano-things with existing communication networks via high-speed Internet. These network paradigms have high application potential in different fields such as healthcare, biomedical, military, smart environment, industry, energy, and multimedia \cite{akyildiz2015internet,cruz2019understanding}. 
Therefore, there is a need for up-to-date studies that comprehensively explain all these network paradigms.

Although there are many studies in IoNT, IoBNT and nano network communication in the literature, there is no survey paper that include Internet of Biodegradable Things (IoBDT) and Internet of Ingestible Things (IoIT). The scopes of existing literature studies in this field are presented in Table \ref{existing_survey}.
Based on this motivation, in this paper, we present a survey that fill the existing gaps in the literature. The main contributions of this survey are summarized as follows: 
\begin{itemize}
    \item We present a recent survey paper covering emerging network paradigms, some of which have been partially explained in the literature (IoNT and IoBNT) and some have not been explained (IoBDT, and IoIT).
    \item We provide the potential application areas, architectural models and communication structures of the all presented network paradigms.
    \item We also highlight different challenges, open issues, and future research directions about all discussed nano-network concepts.
\end{itemize}
The rest of this paper is structured as follows. In Section \ref{iont_iobnt}, we provide an overview of IoNT, IoBNT and their applications. In Section \ref{IoE_BD_T}, we present IoBDT, IoIT and their applications. In Section \ref{challenges}, we discuss challenges, open issues, and future research directions in IoNT, IoBNT, IoBDT and IoIT. Finally, we conclude the paper in Section \ref{conc}.

\section{Internet of Nano Thing and Internet of Bio-Nano Things}
\label{iont_iobnt}
The interconnection of nano-scale devices and machines via the Internet has paved the way for the emergence of new network architectures such as IoNT and IoBNT. These network architectures are based on existing technologies such as IoT, sensor networks, edge, fog, cloud computing and newly developed nano-scale sensor, machine and smart antenna technologies \cite{atlam2018internet}. In this context, the concept of IoNT was described as \textit{“The interconnection of nanoscale devices with existing communication networks and ultimately the Internet defines a new networking paradigm that is further referred to as the Internet of Nano- Things”} in 2010 by Ian F. Akyildiz and Josep Miguel Jornet from the Georgia Institute of Technology \cite{akyildiz2010internet}. On the other hand, IoBNT is a developed version from IoNT and focuses on information exchange, interaction and networking within the biochemical field using synthetic biology and nanotechnology tools \cite{akyildiz2015internet}. The main purpose of IoBNT is to communicate with cells that enable real-time and accurate detection and control of complex biological dynamics occurring in the human body \cite{akyildiz2020panacea}. In other words, while IoNT is based on the integration of nanoscale devices with existing network and communication technologies, IoBNT is additionally based on the behavior and properties of in vitro environments such as molecular communication, Ca$^{2+}$ communication, hormonal communication \cite{bi2021survey}.
The nanodevices used in these network architectures can be created from materials such as biological materials, magnetic fragments or gold nanoparticles. Here, biological devices are created by reprogramming many biological materials such as cells, viruses, bacteria, bacterial phages, erythrocytes \cite{zafar2021systematic}. 
\par In recent years, scientific and technological developments in biomaterial fields  have enabled the development of smart biomaterials with high biofunctions. The smartness level of these biomaterials is divided into four classes as inert, active, responsive and autonomous. These smart materials require a variety of internal and external stimuli to reveal their biofunctionality. Examples of internal stimuli are proteins, enzymes, molecules, antigens, and ionic factor. Examples of external factors are magnetic field, electric fields, light, temperature and mechanical stress \cite{montoya2021road}. 

In the light of the information given above, nano and bio-nano device-based network architectures have many potential application areas such as health, military, environmental monitoring, multimedia and entertainment. Therefore, we first present the network architectures and communication models of IoNT and IoBNT in Subsection \ref{NetArch_Com}. We then provide comprehensive examples of application areas in Subsection \ref{Iont_App_area}.

\subsection{Network Architecture and Communication in IoNT and IoBNT}
\label{NetArch_Com}
\begin{figure*}[!htb]
	\centering
	\includegraphics[width=0.9\textwidth]{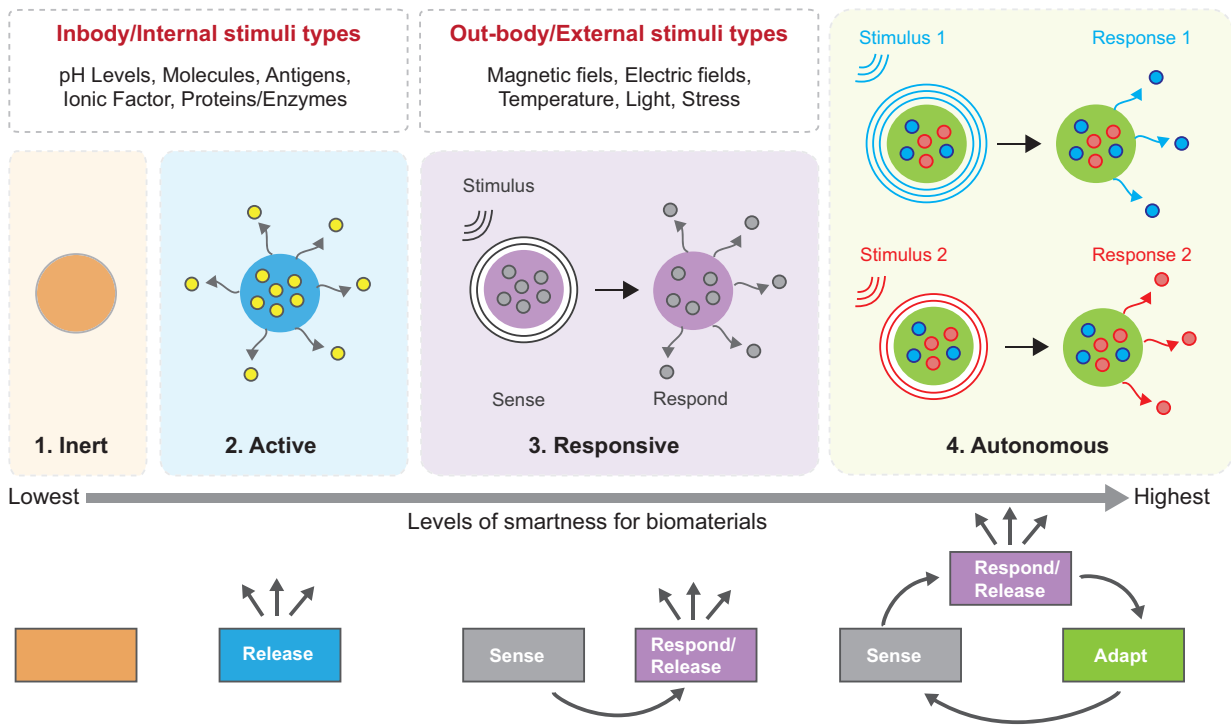}
	\caption{The smartness levels of biomaterials and different internal and external stimuli that enable the biofunctionalities of them. This figure was redrawn by inspiring from \cite{montoya2021road}.}
	\label{bio}
\end{figure*}

Effective integration and communication between nanodevices and macroscale components is required for the IoNT and IoBNT network paradigms to be fully operational. For this reason, the design of network architectures that include several alternative communication paradigms, such as electromagnetic, acoustic, mechanical and molecular communication, comes to the fore \cite{agoulmine2012enabling}. Within these networks, nanothings are expected to interact with each other by exchanging various types of information such as synchronization signals, sensed chemical/physical parameter values, logical operation results, instruction sets, and commands \cite{akyildiz2015internet}.

For the time being, communication in IoNT and IoBNT networks is mainly envisaged as molecular communication and nano-electromagnetic communication \cite{zafar2021securing}. Since these two network paradigms are intertwined, similar technologies can be used in terms of communication. Molecular communication is created by releasing and reacting to certain molecules to transmit or receive information. on the other hand, nanoelectromagnetic communication involves the transmission and reception of an electromagnetic radio frequency waves in the terahertz (THz) band \cite{ali2015internet}. However, there are difficulties in using these communication technologies such as coverage, compatibility, energy and transferred data rate. 
Although there are strong theoretical knowledge and practical models in the literature on communication today, communication at the nanoscale is quite complex and difficult due to the biocompatibility problems of nano devices and the signal propagation properties of tissues in the body \cite{agoulmine2012enabling}.

\begin{table}[!h]
\centering
\caption{Features of communication modes at the nanoscale}
\label{tab:com_modes}
\resizebox{0.45\textwidth}{!}{%
\begin{tabular}{@{}llll@{}}
\toprule
Approach & Biocompatibility & Range in body & Energy \\ \midrule
Nano wireless & Medium to low & 1 cm to 1 m & High \\
Nano acoustic & Low & 1 cm to 10 cm & High \\
Molecular & High to medium & 1 nm to 1 cm & Very low \\
Nanomechanic & Medium to low & 1 mm to 1 nm & Very low \\
Bacteria-based & High to medium & 1 mm to 1 cm & Very low \\ \bottomrule
\end{tabular}%
}
\end{table}

\begin{figure*}[!htb]
	\centering
	\includegraphics[width=1.0\textwidth]{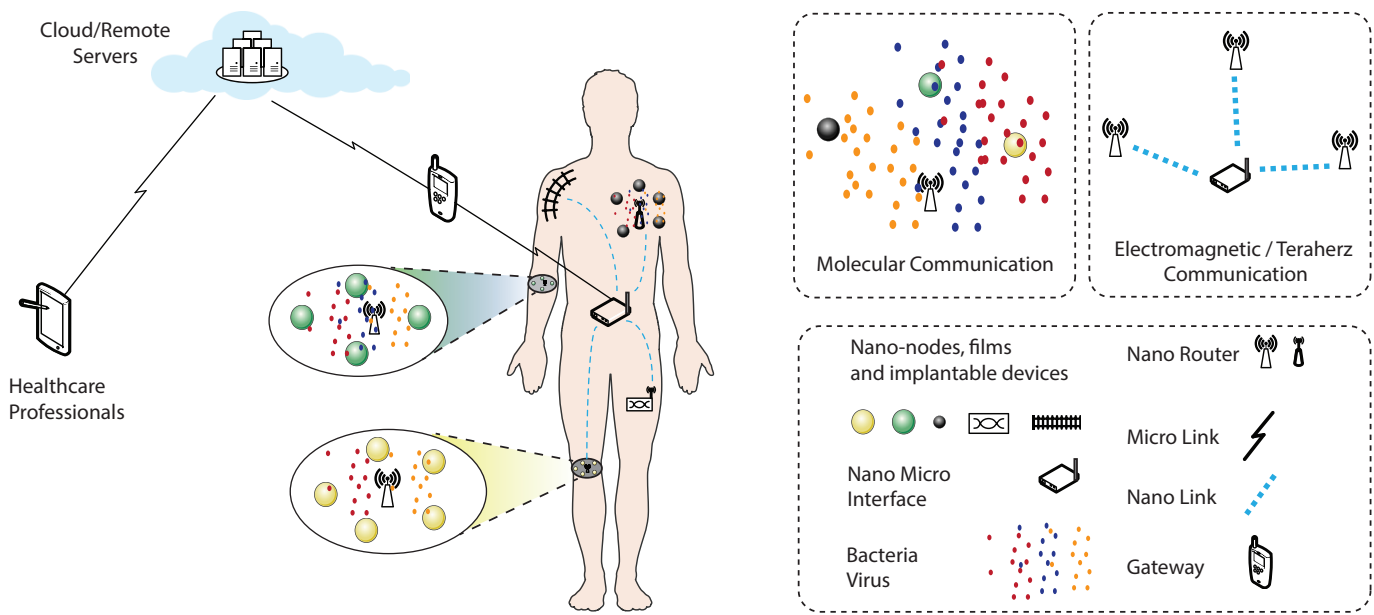}
	\caption{A typical IoNT and IoBNT network architecture for healthcare applications}
	\label{nano_model}
\end{figure*}

IonT and IoBNT network architectures consist of a series of components that connect the electrical field and the biochemical field with devices at the nano or bionano scale. More spesifically, the main components of the IoNT architecture are nano-nodes, nano-routers, nano-micro interface devices and gateways \cite{nayyar2017internet}. IoBNT, on the other hand, may contain additional components that have the function of reading and transferring biochemical domain information. In this context, IoBNT architecture can consist of Bio-cyber interface, gateway and application-specific server components \cite{zafar2021systematic}. All these components are described in Table \ref{tab:components}. Also, a typical IoNT and IoBNT network architecture is shown in Fig. \ref{nano_model}. 

\begin{table*}[!htb]
\caption{The components of the IoNT and IoBNT network architectures}
\centering
\label{tab:components}
\resizebox{0.8\textwidth}{!}{%
\begin{tabular}{@{}ll@{}}
\toprule
Components & Description \\ \midrule
Nano-nodes & \begin{tabular}[c]{@{}l@{}}Nano nodes are the simplest nano devices that perform tasks such as data sensing, \\ transmission, and computation.\end{tabular} \\
Nano-routers & \begin{tabular}[c]{@{}l@{}}Nano routers are more advanced devices than nano-nodes in terms of features \\ such as computing and storage. They are responsible for collecting information \\ from nano nodes and controlling nano nodes with simple control commands \\ (on/off, sleep, read value, etc.).\end{tabular} \\
\begin{tabular}[c]{@{}l@{}}Nano-micro \\ interface device\end{tabular} & \begin{tabular}[c]{@{}l@{}}Nano-micro interfaces are hybrid devices that can both communicate at nanoscale \\ and use classical communication paradigms in traditional communication networks. \\ They are responsible for collecting information from nano routers and sending it to \\ micro-scale devices.  This component is used in IoNT.\end{tabular} \\
Bio-cyber interface & \begin{tabular}[c]{@{}l@{}}The bio-cyber interface is a hybrid device that converts the biochemical \\ signal received from in-body nano-networks into electrical signal for processing \\ by external networks. This component is used in IoBNT.\end{tabular} \\
Gateways & \begin{tabular}[c]{@{}l@{}}Gateways are hybrid devices that can be used in both classical and nano networks \\ at micro and macro scales. These devices allow remote control of the designed IonT \\ and IoBNT networks via the Internet.\end{tabular} \\
\begin{tabular}[c]{@{}l@{}}Application-specific \\ servers\end{tabular} & \begin{tabular}[c]{@{}l@{}}These devices are responsible for the storage, analysis, real-time monitoring of \\ information from nano-networks. They can be used in applications such as \\ healthcare, medical, entertainment or multimedia.\end{tabular} \\ \bottomrule
\end{tabular}%
}
\end{table*}

\subsection{IoNT and IoBNT Applications}
\label{Iont_App_area}
IoNT and IoBNT network paradigms have added new dimensions to the existing application areas of IoT such as healthcare and communication networking. Therefore, in this section, we present current studies on these two application areas. 

In healthcare, it has been shown that these network paradigms  can be used in the treatment of circulatory system diseases, infectious Diseases, diabetes, thrombosis and many physical or psychological diseases. Today, due to the Covid-19 pandemic, remote work has become increasingly common in many areas. In this context, studies have been carried out regarding the electronic and remote provision of health services \cite{zafar2021systematic}. In this way, it has become possible to transmit data collected from patients connected to the IoNBT network directly and in real time to healthcare professionals. Thanks to the remote collection of patient data, the patient may not need to go to the laboratory for testing. In addition, in case of any infection, the disease can be detected even before the patient shows symptoms and medical support can be given to the patient \cite{akyildiz2020panacea}. Telemedicine applications related to these sample scenarios have been included in our lives with the covid 19 epidemic. For example, Jarmakiewicz et al. set forth standards and guidelines for the research and development of nanosensor networks for some telemedicine applications. They also demonstrated that some applications can be implemented with a nano-sensor network by testing it with a nano-network developed for human circulatory system \cite{jarmakiewicz2016internet}. 

Thrombosis is one of the leading causes of death worldwide, killing one in four people. In thrombosis, a blood clot forms in a vein, once the blood clot forms, it can stop or delay blood flow, or it can be found in organs. Froud et al. \cite{fouad2020analyzing} created a new IoBNT-based model with a bio-cyber communication interface that helps predict and analyze blood vessel coagulation. Thanks to this model, the information in the blood vessel is collected and the bio-cyber interface is used to convert the information into electrical equivalent. Moreover, optical or thermal response has also been used to stimulate the release of certain nanocarrier molecules such as liposomes, nanodevices that can be transported through the bloodstream and predict clots.

IoBNT can also be used in the early diagnosis and  mitigation of infectious diseases. For instance, cystic fibrosis disease is a genetic disease that can be seen in organs and systems such as lungs, pancreas, intestines, sweat glands. It develops in waves and causes the death of patients. In \cite{akyildiz2020panacea}, Akyıldız et al. proposed an IoBNT network called PANACEA, which provides an end-to-end solution to infectious diseases. In this network, a submillimeter implantable bio-electronic device that senses communication within body cells is used to determine the level of infection. 

Another common disease today is diabetes. Abbasi et al. \cite{abbasi2017information} focused on modeling IoBNT applications that will improve the diagnosis, management and treatment practices of the insulin-glucose system for this disease. Through an IoBNT network to be developed, insulin and glucose concentration values can be transmitted directly to healthcare providers. In this way, besides monitoring the health status, pump life can be extended by ensuring that the insulin cartridge lasts longer. In addition, thanks to such smart systems, the body is protected from the side effects of excess insulin.
On the other hand, preventive health services can be provided by methods such as pre-illness therapy with a holistic approach to physical and psychological diseases. In this context, it is predicted that healthy living conditions can be created by making changes in people's lifestyles with IoBNT \cite{sarker2019holistic}.

Since IoNT and IoBNT are emerging network paradigms, the developed applications in these networks are also at the beginner level. Moreover, the integration of nanomachines into the human body, their architectural design, and efficient operation (communication, computing, storage,  etc.) of these devices within the heterogeneous network structure are among the main challenges. Therefore, there is an increasing number of research efforts aimed at eliminating existing and potential challenges. Al-Turjman proposed an energy efficient framework focusing on data delivery in nanonetworks. With this framework, it is aimed to realize data distribution with energy sensitive routing protocols by considering the shortest path \cite{al2020cognitive}.

It is clear that IoNT has different architectural requirements for different network models and applications. In these networks, communication model design is also another challenge. In \cite{ali2016internet}, Ali et al. investigated the structure of communication models in the IoNT network that developed for drug delivery and disease detection. In the study, they evaluated the advantages and disadvantages of these two models by establishing non-additive and single-layer communication models.
In \cite{stelzner2017function}, Stelzner et al. introduced a new concept of function-centered Nano-network (FCNN) focusing on intra-body communication scenarios. FCNN aims to minimize memory requirements by combining the location and function capabilities of nano machines. 
In another study, Canovas-Carrasco et al. \cite{canovas2019optimal} emphasized that the development of optimal transmission policies to be used in nano-networks, reducing implementation costs and in vivo monitoring of nano-sensors depend on the generation of smart policies. For this reason, the authors proposed the Markov decision process model, which enables the derivation of smart policies.

From the network architecture perspective,, Galal and Hesselbach \cite{galal2018nano} presented a multi-layered architectural model in nano networking that combines software-defined networking (SDN), network function virtualization (NFV) and IoT technologies. In the study, the authors proposed a number of functionalities and usage scenarios that could be implemented for nanodevices. In addition, significant challenges and gaps in implementing the proposed functions with nano-technology are discussed.
In \cite{pramanik2020advancing}, the usage possibilities, architecture, communication models, advantages and challenges of nano technology through nanobiosensors and IoNT in modern health care were determined. It has been emphasized that the concept of placing nanoscale devices in the human body has the potential to be used in all health and medical applications that exist today.

\section{Internet of Biodegradable and Ingestible Things}
\label{IoE_BD_T}
Today, the world is faced with a serious waste load due to global warming, climate change and environmental pollution. Therefore, the biological and technological needs for the correct disposal of the generated waste are increasing day by day \cite{velvizhi2020biodegradable}. The ability to recycle biodegradable, compostable and soluble materials to eliminate the negative effects attracts the attention of scientists and the public more than ever before. 
Recent advances in synthetic chemistry and engineering have enabled the development of many materials such as medical implants, films, environmental sensors, non-injectable sensors, edible and biodegradable sensors, films, and disposable devices \cite{salvatore2017biodegradable,ashfaq2022gelatin}. Biodegradable and editable electronics work for a period of time and are then lost by hydrolysis or biochemical reactions. Intelligent forms of these devices can be used in a wide variety of fields such as food packaging, drug delivery, tissue engineering, and medical equipment manufacturing \cite{salvatore2017biodegradable}. The structure of biodegradable devices/sensors can compose of metals, polymers, silicon and their composites \cite{ashammakhi2021biodegradable}. Moreover, edible and digestible devices can be swallowed by patients daily in pill form or consumed with food. In addition to being suitable for ingestion, edible devices are fully digested in the body and can be safely released into the environment. Thanks to these features, the device can transmit patient health data to healthcare professionals in real time. In addition, it can eliminate the complications (bleeding, perforation of the colon, severe pain in the abdomen) that may arise from routine scanning and imaging techniques such as endoscopy and colonoscopy \cite{steiger2019ingestible}.

\begin{figure*}[!htb]
	\centering
	\includegraphics[width=1.0\textwidth]{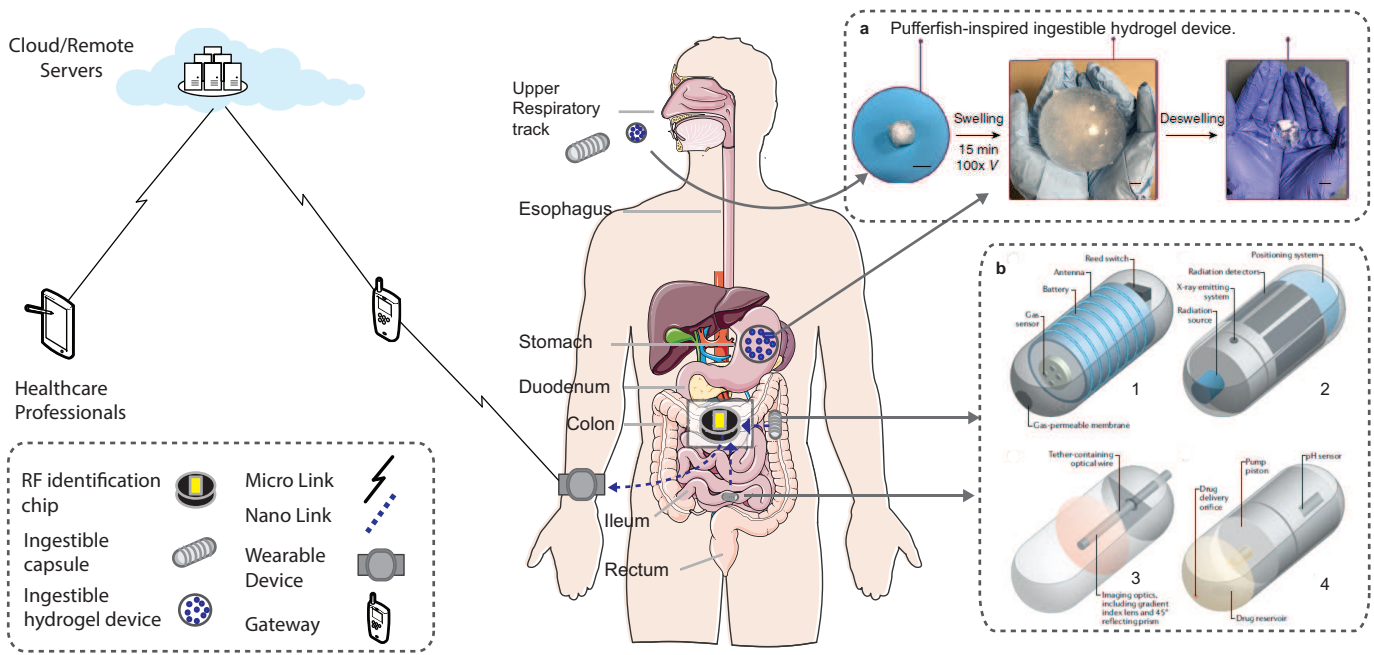}
	\caption{A typical IoBDT and IoIT network architecture. a) Pufferfish-inspired ingestible hydrogel device \cite{liu2019ingestible}, b) Ingestible electronics/capsules \cite{steiger2019ingestible} - 1. Gas-sensing capsule \cite{kalantar2018human}, 2. X- ray scanning capsule, 3. Optical coherence tomography \cite{gora2013tethered}, 4. IntelliCap for drug delivery \cite{becker2014novel}. The human body and digestive system organs used in this Figure are taken from Servier Medical Art \cite{ServierMedicalArt}, which is allowed free access and use.}
	\label{ioit_model}.
\end{figure*}

Thanks to the advances in biochemistry, materials and engineering, it has become possible to use the above-mentioned different forms of electronic and biomaterials for the benefit of the environment and human beings. At this point, IoBDT and IoIT networks emerge as specific types of IoBNT. These networks can be developed based on IoNT and IoBNT specific components and architectures. However, these networks differ in terms of edge device/sensor and consist of application-specific edible, digestible and biodegradable sensors or devices. An example system/network model of IoBDT and IoIT is given in Fig \ref{ioit_model}.
\subsection{IoBDT and IoIT Applications}
It is seen that degradable, digestible and edible devices are widely used in the digestive and circulatory systems. In this context, in \cite{gonzalez2007ingestible} the authors stated that edible electronic devices can be used to monitor the physical and chemical parameters of the gastrointestinal system such as pH, temperature and peritalsis, as well as various potagens and disorders in the gastrointestinal system. In another study, Steiger et al. \cite{steiger2019ingestible} showed that capsule endoscopes can be used to detect bleeding in the gastrointestinal tract and detect inflammatory bowel diseases. In \cite{philip2017review}, an example IoIT network design is presented, where information about the patient's stomach can be transmitted to doctors over the cloud via swallowable tablets.
In the study \cite{becker2014novel}, Becker et al. explained that edible smart pills can be used for digital monitoring of chronic patients' drug intake, targeted drug delivery, drug release rate, drug intake time and place.

On the other hand, various studies have been conducted to reveal the state of intestinal health and disorders, the effect of foods, medicinal supplements and environmental changes on the gastrointestinal tract. The main purpose of these studies is to detect and monitor intestinal structure and functions using ingestible sensing capsule technology \cite{kalantar2017ingestible}.

In \cite{kalantar2018human}, Kalantar-Zadeh et al. conducted a pilot trial in a real human body of an ingestible electronic capsule containing thermally conductive and semiconductor sensors that detect oxygen, hydrogen, and carbon dioxide gases. The authors inferred intestinal microbial fermentative activities and gas profiles by varying the subjects' dietary fiber intake.

However, the produced electronic sensors must have a fast response capability, and their biodegradation and waste conditions must be controlled after use. For example, ingestible chemical sensors record changes and pH values in the gastrointestinal tract in the detection of ulcers and stomach tumors \cite{sahoo2020smart}. In addition, oral medications are exposed to a wide pH range. For this reason, drug components may degrade at the wrong times and their effect may decrease. However, drugs are released with environmental sensors, feedback algorithm and drug release procedures in smart pills \cite{sahoo2020smart}.
To this end, In \cite{salvatore2017biodegradable}, Salvatore et al. are focused on producing a temperature sensor with ultra-thin format, fast response time (10ms) and stable operation with little resistance change when the device is folded At the end of the study, the authors developed a sensor that provides high mechanical stability and is also biodegradable.
In another study, Liu et al. \cite{liu2017degradable} introduced a ingestible hydrogel device that can be swallowed as a pill, has a rapid transition feature, and can take the form of a soft sphere. This Ingestible hydrogel device represents a non-invasive device that swells in the stomach within 60 minutes, remains intact in the stomach cavity. It has the potential to interact closely with the digestive system in the human body, thanks to its in vitro drug release and in vivo temperature sensing applications. The authors tested the indigestible hydrogel device for long-term gastric retention and physiological monitoring in large animal groups.

\section{Challenges, Open Issues, and Future Research Directions }
\label{challenges}
IoNT, IoBNT, IoBDT and IoIT networks have wide application potential to improve the health and quality of life of living things. However, many challenges are encountered in real-life applications of these networks. These challenges also present the topic of open research issues and future research directions.

\begin{table*}[!htb]
\caption{Summary of some selected papers in the area of IoNT, IoBNT, IoBDT and IoIT}
\resizebox{\textwidth}{!}{%
\begin{tabular}{@{}llllll@{}}
\toprule
Paper \hphantom{00}               & Scope                                                        & Domain                                                                                    & Technology                                                                & \begin{tabular}[c]{@{}l@{}}Obtained/Used\\  Datasets\hphantom{00}\end{tabular} & Contribution of the work                                                                                                                                                                                                                                                                                                                                                                          \\ \midrule
\cite{stelzner2017function}       & IoNT                                                         & \begin{tabular}[c]{@{}l@{}}Biomedical \\ (Human Body)\end{tabular}                        & \begin{tabular}[c]{@{}l@{}}Sensor \\ Actuator\end{tabular}                & None                                                                           & \begin{tabular}[c]{@{}l@{}}They tried to minimize memory requirements by \\ combining the data collected about the location \\ and function of the respective nano-devices, using \\ the Function-centered Nano-network(FCNN) \\ architecture.\end{tabular}                                                                                                                                       \\
\cite{jarmakiewicz2016internet}   & IoNT                                                         & Healthcare                                                                                & \begin{tabular}[c]{@{}l@{}}Nanosensor \\ Radio channels\end{tabular}      & Simulation Data                                                                & \begin{tabular}[c]{@{}l@{}}The telemedicine applications in the literature were \\ examined and the points that could be improved \\ were determined. It has been stated that it would be \\ beneficial to support these applications by using a \\ nano-sensor network that has the potential to be \\ applied in some scenarios and that works in the \\ human circulatory system.\end{tabular} \\
\cite{galal2018nano}              & Nano Networks\hphantom{0}                                                & Healthcare                                                                                & \begin{tabular}[c]{@{}l@{}}Nano devices\\ Nano-antennas\end{tabular}      & \begin{tabular}[c]{@{}l@{}}Nano drugs\\ Wearable devices\end{tabular}          & \begin{tabular}[c]{@{}l@{}}It is represented an approach which is combining \\ software-defined networking (SDN), network \\ function virtualization (NFV), and IoT. Also,\\ a composite architecture model of nano network \\ communication is proposed.\end{tabular}                                                                                                                            \\
\cite{canovas2019optimal}         & \begin{tabular}[c]{@{}l@{}}IoNT\\ Nano Networks\hphantom{0}\end{tabular} & \begin{tabular}[c]{@{}l@{}}Healthcare \\ (Human \\ Cardiovascular \\ System)\end{tabular} & \begin{tabular}[c]{@{}l@{}}Nano devices\\ Nano-routers\end{tabular}       & \begin{tabular}[c]{@{}l@{}}Cardiovascular \\ system\end{tabular}               & \begin{tabular}[c]{@{}l@{}}A Markov-based decision process model is proposed \\ to generate optimal transmission policies to be used\\ by nano nodes. In addition, a series of simulations \\ were executed.\end{tabular}                                                                                                                                                                         \\
\cite{salvatore2017biodegradable} & IoBDT / IoIT                                                        & Food Science                                                                              & \begin{tabular}[c]{@{}l@{}}Bluetooth\\ Nanomaterials\end{tabular}         & None                                                                           & \begin{tabular}[c]{@{}l@{}}It has been developed a sensor that can operate at high \\ temperatures by using nanofilm. This sensor is tested on \\ the food tracking system.\end{tabular}                                                                                                                                                                                                          \\
\cite{kalantar2018human}          & IoBDT / IoIT                                                        & \begin{tabular}[c]{@{}l@{}}Healtcare\\ (Human \\ Digestive \\ System)\end{tabular}        & \begin{tabular}[c]{@{}l@{}}Gas sensors\\ Electronic capsules\end{tabular} & None                                                                           & \begin{tabular}[c]{@{}l@{}}An ingestible electronic capsule is used to follow the \\ effects on the habits of an individual's diet,  this \\ electronic capsule (containing different sensors) \\ is tested in the human body (intestinal flora).\end{tabular}                                                                                                                                    \\
\cite{liu2019ingestible}          & IoBDT / IoIT                                                        & \begin{tabular}[c]{@{}l@{}}Healtcare \\ (Gastrointestinal \\ Tract)\end{tabular}          & Hydrogel Device                                                           & \cite{hdd}                                                                     & \begin{tabular}[c]{@{}l@{}}A Hydrogel device has been produced that can be \\ swallowed like a pill and provides data transfer.\end{tabular}                                                                                                                                                                                                                                                      \\ \bottomrule
\end{tabular}%
}
\end{table*}

\begin{itemize}
    \item Nano-network architecture: Traditional IoT networks are usually built with three, five and seven layers. In nano-networks, it is very difficult to establish a layered network structure in terms of device size, computation, storage capacity and communication types.
    \item Nano communication and Standards: There are short and limited (from 1 nm to 1 m) in-vivo communication modes in nano networks. There is also a high level of biological noise in the environment. These communication difficulties can cause data loss, delay, and network congestion in nano communication. In addition, since these networks differ from conventional networks in many aspects such as operating environment and communication mode, the OSI/TCP model is not fully suitable for these networks. In this context, the foundation of new communication standards is quite necessary \cite{kuscu2021fabrication,pramanik2020advancing}.
    \item Bio-cyber Interface: As mentioned earlier, different bio-cyber interfaces have been proposed for each considered network. However, these interfaces are not yet at the practical and clinically desirable level. For these reasons generalized and realistic biointerface deployments are needed \cite{bakhshi2019securing}.
    \item Data Management and Analysis: It is predicted that there will be a rapid increase in the number of new devices connected to the internet together with nano networks. Currently, data method and analysis is a serious challenge. This challenge will become a hot topic with nanonetworks. In addition, the development of smart data analysis algorithm models that can work on nano-scale devices is open to research \cite{balasubramaniam2012realizing}.
    \item Smart Biomaterials: Although there are many types of biocompatible in-capsular sensors (camera, pH, temperature, gas, ultrasound imaging, physisorption, surface acustic wave, and etc.), the field of ingestible sensors and smart biomaterials is still in its absolute infancy \cite{kalantar2017ingestible,montoya2021road}.
    \item Security and Privacy: The new nano-network paradigms mostly include applications that have not yet established standards and frameworks in the medical, biological, chemical or personal fields. Therefore, these networks are currently vulnerable to a very large threat surface with multidimensional attack vectors \cite{ali2021emerging}. At this point, ensuring the confidentiality, integrity and availability of the data stored and circulating on the network is more critical than other application areas. The weakness that will arise in this regard may harm individuals and societies by laying the groundwork for negative situations such as data manipulation, theft, espionage and even bioterrorism \cite{giaretta2015security}. Therefore, trust management systems, user access control systems, lightweight data encryption and compression models are needed.
    \item Other challenges: In addition to the existing challenges, there are many uncovered research topics such as content management, mobility management, service aggregation and discovery, energy conservation, energy harvesting, power transfer, nano device addressing, integration with next generation technologies (SDN, NFV, Blockchain, NFT, Metaverse and etc.) \cite{balasubramaniam2012realizing,kuscu2021internet}.
\end{itemize}
\section{Conclusion}
\label{conc}
This paper provides a comprehensive literature to contribute to a holistic understanding of emerging nano-networks. For this purpose, we present all of the promising network paradigms such as IoNT, IoBNT, IoBDT and IoIT that have emerged based on nanoscale devices and materials in connection with "All-thing connected" vision. We give comprehensive explanations of communication, network architecture and application domains of these network paradigms. Moreover, we comprehensively provide the challenges of these networks, open issues, and future research guidelines. In this work, we comprehensively present all emerging nanonetwork paradigms in a single survey paper, and we hope that the paper will shed light on future works.

\section*{Acknowledgement}
Seyda Sentürk is supported by the Council of Higher Education (CoHE) under the special 100/2000 scholarship program.

\bibliographystyle{IEEEtran}
\bibliography{reference}
\end{document}